\documentclass[usenatbib]{mn2e}
\usepackage{graphicx,tabulary,amsmath,amssymb,upgreek,wasysym,subfigure,overpic,array}
\usepackage[T1]{fontenc}
\usepackage{upgreek}

\newcommand{\changes}[1]{{#1}}

\numberwithin{equation}{section}
\title[]{Forming isolated brown dwarfs by turbulent fragmentation}
\author[O. Lomax, A. P. Whitworth, D. A. Hubber]{O. Lomax\thanks{E-mail: oliver.lomax@astro.cf.ac.uk}$^1$, A. P. Whitworth$^1$, D. A. Hubber$^{2,3}$\\
$^1$School of Physics and Astronomy, Cardiff University, Cardiff CF24 3AA, UK\\
$^2$University Observatory, Ludwig-Maximilian-University Munich, Scheinerstr.1, D-81679 Munich, Germany\\
$^3$Excellence Cluster Universe, Boltzmannstr. 2, D-85748 Garching, Germany}

\begin{document}
\pagerange{\pageref{firstpage}--\pageref{lastpage}} \pubyear{2015}
\maketitle
\label{firstpage}

\begin{abstract}
We use Smoothed Particle Hydrodynamics to explore the circumstances under which an isolated very-low-mass prestellar core can be formed by colliding turbulent flows and collapse to form a brown-dwarf. Our simulations suggest that the flows need not be very fast, but do need to be very strongly convergent, i.e. the gas must flow in at comparable speeds {\it from all sides}, which seems rather unlikely. We therefore revisit the object Oph-B11, which \citet{AWG12} have identified as a prestellar core with mass between $\sim 0.020\,\mathrm{M_{_\odot}}$ and $\sim 0.030\,\mathrm{M_{_\odot}}$. We reanalyse the observations using a Markov-chain Monte Carlo method that allows us (i) to include the uncertainties on the distance, temperature and dust mass opacity, and  (ii) to consider different Bayesian prior distributions of the mass. We estimate that the posterior probability that Oph-B11 has a mass below the hydrogen burning limit at $\sim 0.075\,\mathrm{M_{_\odot}}$, is between 0.66 and 0.86\,. We conclude that, if Oph-B11 is destined to collapse, it probably will form a brown dwarf. However, the flows required to trigger this appear to be so contrived that it is difficult to envisage this being the only way, or even a major way, of forming isolated brown dwarfs. Moreover, Oph-B11 could easily be a transient, bouncing, prolate core, seen end-on; there could, indeed should, be many such objects masquerading as very low-mass prestellar cores. 
\end{abstract}


\section{Introduction}\label{SEC:INTRO}

Brown dwarfs are very low-mass stars which, as they condense out of the interstellar medium, become sufficiently dense to be supported by electron degeneracy pressure before they become hot enough to start burning hydrogen. Their existence was predicted theoretically by \citet{K63} and \citet{HN63}. They were first observed thirty years later, by \citet{RZM95} and \citet{NOK95}. They have masses $\la 0.075\,\mathrm{M_{_\odot}}$, and it is estimated \citep[e.g.][]{AMGA08} that there is roughly one brown dwarf for every four hydrogen-burning stars. The principal mechanism by which brown dwarfs form is unclear.

 \citet{RC01} have suggested that brown dwarfs might be formed when a core of intermediate or high mass spawns a small cluster of stars; dynamical interactions between the stellar embryos result in some of them being ejected before they have acquired sufficient mass to be hydrogen-burning stars, and once ejected they are unlikely to grow further, because their environment is too rarefied and their velocity too high. However, there is little evidence for the diaspora of high-velocity brown dwarfs around young clusters that this mechanism would produce.

\citet{HSS96} have pointed out that brown dwarfs could be formed when prestellar cores are overrun by H{\sc ii} regions and photo-evaporated. The inner parts of the core collapse, but the outer layers are boiled away, and consequently the resulting star has very low mass. However, \citet{WZ04} calculate that this mechanism is extremely inefficient, in the sense that it requires quite a massive core to form a single brown dwarf. Furthermore, it can only work in clusters where there are massive ionising stars, so it is unlikely to be the dominant formation mechanism for brown dwarfs, since these are observed to be abundant in star formation regions like Taurus, where there are no ionising stars.

Another possibility is that brown dwarfs form by the gravitational fragmentation of extended massive accretion discs around larger protostars. Detailed radiation-hydrodynamic simulations of core collapse and fragmentation, starting from initial conditions informed as closely as possible by state-of-the-art observations, indicate that brown dwarfs can form, with the correct distributions of mass, multiplicity and multiplicity statistics --- but only if the initial turbulent velocity field has a significant solenoidal component, {\it and} if accretion onto the primary protostar at the centre of the accretion disc (and hence also radiative feedback from this protostar) is episodic, with a duty cycle $\ga 3000\,{\rm yr}$ \citep{SHW07,SW08,SW09a,SW09b,TKGSW,SMWA11,LWHSW14,LWHSW14b,LWH15}. Moreover, these simulations do not include magnetic fields, and it is unclear whether sufficiently extended massive accretion discs can form when the angular momentum transport facilitated by magnetic fields is included \citep{BH98,JHC12,JHCF13}.

We stress that none of the above three mechanisms should apply exclusively to brown dwarfs, as distinct from hydrogen-burning stars. Rather it is anticipated that, as one proceeds to less massive stellar populations, across the hydrogen-burning limit, these mechanisms may be responsible for delivering an increasing fraction of the stars making up those populations.

A fourth possibility is that brown dwarfs form by turbulent fragmentation \citep[e.g.][]{PN02,HC08,HC09}. In turbulent fragmentation, star formation occurs wherever colliding turbulent streams produce a condensation that is Jeans unstable, in other words a core that is sufficiently dense, cool and quiescent to be prestellar. This prestellar core then collapses and possibly also fragments, to form a star or stars. This is how the majority of H-burning stars are presumed to form, and indeed how the embryos invoked in the ejection theory of \citet{RC01} are presumed to form, and how the accretion discs invoked in the simulations of \citet{LWHSW14,LWHSW14b} and \citet{LWH15} do form. An additional consideration is that turbulent fragmentation involves the formation of filaments, and these can fragment to produce low-mass cores; in large-scale simulations of whole molecular clouds\citep[e.g.][]{B09b,B12} many brown dwarfs appear to form in this way, i.e. not in isolation, but as part of a filamentary structure. 

If brown dwarfs form in the same way as Sun-like stars, then, since the binary frequency for brown dwarfs is low \citep{CSFB03,BR06}, brown dwarfs should be able to form in isolation, and this possibility has recently recieved observational support with the discovery of an object, Oph-B11, that appears to be an isolated prestellar core of brown-dwarf mass \citep{AWG12}. However, the ram pressures required to create a Jeans-unstable core of brown-dwarf mass are very large, and the shocks very strong, so that the gas is heated to high temperature, and will tend to squirt out in directions parallel to the shock, unless there are strong inflows from all directions. It is this requirement that we seek to evaluate in this paper. We conclude that it is a rather limiting requirement, and therefore we revisit the analysis of the observations of Oph-B11, to place somewhat more conservative uncertainties on its mass. We also suggest some alternative explanations for the status of Oph-B11.

In Section \ref{SEC:SIMS}, we present numerical simulations of the formation and collapse of prestellar cores having masses of  $0.020\,\mathrm{M_{_\odot}}$ and $0.060\,\mathrm{M_{_\odot}}$. We conclude that collapse requires relatively modest inflow velocities, but a very high degree of convergence, i.e. gas flowing in from all sides, and therefore that this mechanism can only operate under exceptional circumstances. In Section \ref{SEC:MASS} we revisit the assumptions involved in estimating the mass of Oph-B11, and develop a Bayesian analysis to estimate the probability that it is below the hydrogen-burning limit. We find that, provided Oph-B11 is approximately spherical, and the N$_2$H$^+(1-0)$ detection is valid, there is a $66$ to $86\%$ chance that its mass is below the hydrogen-burning limit. In Section \ref{SEC:ALTS} we explore an alternative explanations for Oph-B11, namely that it may simply be a transient, bouncing, prolate condensation, seen by chance end-on, and therefore displaying a very small line-of-sight non-thermal velocity dispersion. In Section \ref{SEC:CONC} we summarise our conclusions.

\section{Simulations of the formation of Oph-B11}\label{SEC:SIMS}

The majority of observed prestellar cores in Ophiuchus have mean densities $\rho_{_{\rm CORE}}\sim 10^{-17\pm1}\,\mathrm{g\,cm^{-3}}$ \citep[e.g.][]{PWK15}. However, an object with temperature $T\sim 10\,\mathrm{K}$ and $M_\textsc{core}\sim 0.020\,\mathrm{M_\odot}$ is only Jeans unstable if it has a mean density $\rho_{_{\rm CORE}}\gtrsim 5\times 10^{-15}\,\mathrm{g\,cm^{-3}}$. In this section, we report Smoothed Particle Hydrodynamics simulations designed to explore what turbulent flows are required to form an isolated prestellar core (i.e. a Jeans unstable core) with brown-dwarf mass, in particular how fast and how convergent the flows have to be.

\subsection{Initial conditions}\label{SSEC:InitConds}

To set up the initial conditions, we start by position ${\cal N}_{_{\rm CUBE}}=190986\;$ equal-mass SPH particles randomly in a cubic box. Then we relax their positions with periodic boundary conditions, assuming they have a universal temperature and invoking only pressure and artificial viscosity forces, to produce a uniform-density glass. Next we cut out from the cube a sphere with diameter equal to the side of the cube, so it contains ${\cal N}_{_{\rm SPHERE}}\simeq \pi{\cal N}_{_{\rm CUBE}}/6\simeq 100000$ particles. Then we scale the relative positions of the particles so that the radius of the sphere is $R_{_{\rm CORE}}=0.02\,{\rm pc}$, and we scale the masses of the SPH particles to produce two different prestellar cores. One has mass $M_{_{\rm CORE}}=0.02\,{\rm M}_{_\odot}$ (hence its initial volume-density is $\rho_{_{\rm CORE}}=3.2\times 10^{-19}\,{\rm g}\,{\rm cm}^{-3}$, the mass of a single SPH particle is $m_{_{\rm SPH}}=2\times 10^{-7}\,{\rm M}_{_\odot}$, and the mass resolution, corresponding to $\sim\! 100$ SPH particles, is $\;M_{_{\rm MIN}}\sim 2\times 10^{-5}\,{\rm M}_{_\odot}$). The second has mass $M_{_{\rm CORE}}=0.06\,{\rm M}_{_\odot}$ (hence $\rho_{_{\rm CORE}}=9.7\times 10^{-19}\,{\rm g}\,{\rm cm}^{-3}$, $m_{_{\rm SPH}}=6\times 10^{-7}\,{\rm M}_{_\odot}$, and $M_{_{\rm MIN}}\sim 6\times 10^{-5}\,{\rm M}_{_\odot}$).

The initial velocity field is specified with two parameters, $(v_{_{\rm O}},c)$. $v_{_{\rm O}}$ gives the initial velocity magnitude for all SPH particles. $c$ is a convergence parameter which measures the extent to which the velocity field is converged on the centre of the core {\it from all sides}. Specifically, the initial velocity field is given by
\begin{equation}
 \boldsymbol{v}=-\,v_{_{\rm O}}\,\frac{c\,\hat{\boldsymbol{r}}+(1-c)\,\hat{\boldsymbol{z}}}{|c\,\hat{\boldsymbol{r}}+(1-c)\,\hat{\boldsymbol{z}}|}\,.
\end{equation}
Here, $\hat{\boldsymbol{r}}=(x,y,z)/\sqrt{x^2+y^2+z^2}$ is the unit position vector and $\hat{\boldsymbol{z}}=(0,0,z)/\lvert z\rvert$ is the unit $z$-direction vector. The velocity is therefore the $c$-weighted average of a radially convergent flow and two anti-parallel flows. When $c=0$, the system initially comprises two hemispheres colliding head-on. When $c=1$, the system initially converges radially on a single point. Intermediate values of $c$ result in a normalised superposition of these two extremes. Conceptually it is easier to envisage turbulent flows delivering approximately planar shocks $(c\sim 0)$; strongly radially convergent flows $(c\sim 1)$ would seem to be very unlikely.

\begin{figure}
\centering
\includegraphics[width=\columnwidth]{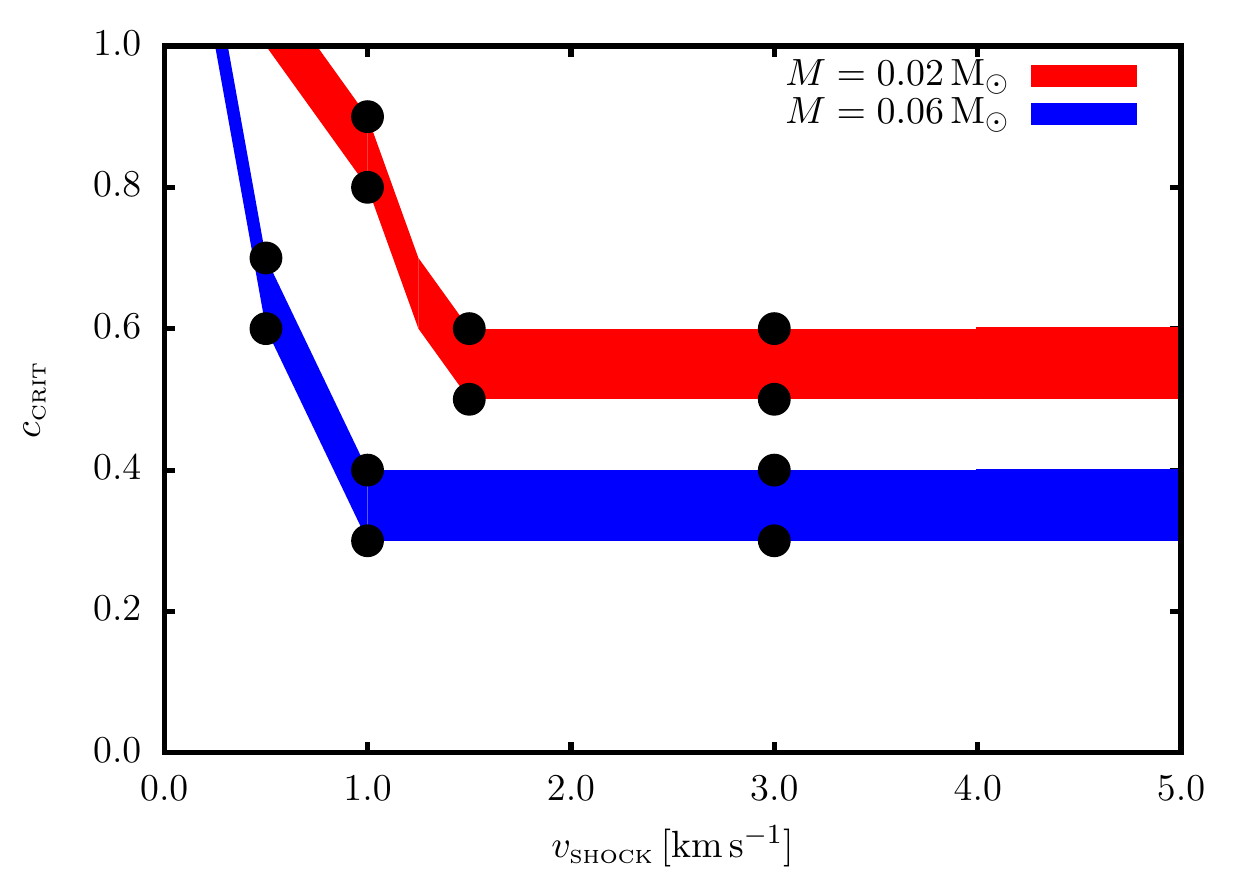}
\caption{The $(v_{_{\rm O}},c)$-plane. Values of $(v_{_{\rm O}},c)$ above and to the right of the red band represent the cases where the colliding flow is sufficiently strong and convergent to trigger the formation of a brown dwarf in a core with mass $0.020\,{\rm M}_{_\odot}$; the width of the band reflects the quantisation of the cases we have simulated. The blue band shows the same information for a core with mass $0.060\,{\rm M}_{_\odot}$. The black filled circles show the cases which were repeated with four times as many SPH particles, to demonstrate convergence.}
\label{fig:c_crit}
\end{figure}

\subsection{Numerical Method}\label{SSEC:NumMeth}

We use the \textsc{seren} $\nabla h$-SPH code \citep{HBMW11} to simulate the evolution of cores. Self-gravity is computed using a tree, and we invoke the \citet{MM97} formulation of time dependent artificial viscosity. The spatial resolution parameter is set to $\eta =1.2$, so an SPH particle typically has 56 neighbours in its smoothing kernel. The opacity limit ($\sim\!\!3\times10^{-3}\,\mathrm{M}_{\odot}$) is resolved with $\sim\!15000$ SPH particles when $M_{_{\rm CORE}}=0.020\,{\rm M}_{_\odot}$, and $\sim 5000$ SPH particles when $M_{_{\rm CORE}}=0.060\,{\rm M}_{_\odot}$. If a gravitationally bound condensation forms at a minimum in the gravitational potential, where the density exceeds $10^{-9}\,{\rm g}\,{\rm cm}^{-3}$, it is replaced with a sink particle \citep{HWW13}. Sink particles have a radius corresponding to the radius of an SPH particle with density equal to $\rho_{_{\rm SINK}}$ (i.e. $0.12\,\mathrm{au}$ when $M_{_{\rm CORE}}=0.020\,{\rm M}_{_\odot}$, and $0.17\,\mathrm{au}$ when $M_{_{\rm CORE}}=0.060\,{\rm M}_{_\odot}$). Sink particles are not allowed to form within the radius of a pre-existing sink particle. The equation of state and the energy equation are treated with the algorithm described in \citet{SWBG07}, which captures the ionisation states of hydrogen and helium and their contributions to the internal energy, as well as the evolution of the mass opacity coefficient (dust sublimation, molecular abundances, etc.), and transport of the cooling radiations emitted by different species. The gas starts off with temperature $T\sim 10\,{\rm K}$, and hence isothermal sound speed $a\sim 0.2\,{\rm km}\,{\rm s}^{-1}$.

\subsection{Results}\label{SSEC:Sim_Results}

\subsubsection{Overview}\label{SSSEC:Overview}

For both core masses, we have simulated the evolution with  
\begin{eqnarray}\nonumber
v_{_{\rm O}}&=&0.25,\,0.5,\,1.0,\,1.5,\,2.0,\,3.0,\,4.0\text{ and }5.0\,\mathrm{km\,s^{-1}}\,,\\\nonumber
c\hspace{0.23cm}&=&0,\,0.1,\,0.2,\,0.3,\,0.4,\,0.5,\,0.6,\,0.7,\,0.8,\,0.9\text{ and }1.0\,.
\end{eqnarray}
As might be expected, formation of a sink, i.e. collapse beyond a density of $\rho_\textsc{sink}=10^{-9}\,\mathrm{g}\,\mathrm{cm}^{-3}$, requires a large $v_{_{\rm O}}$ and/or a large $c$. 

The red band on Fig. \ref{fig:c_crit} separates (i) the smallest combination of $v_{_{\rm O}}$ and $c$ that leads to collapse from (ii) the largest combination that does not lead to collapse, for the core with $M_{_{\rm CORE}}=0.020\,{\rm M}_{_\odot}$. In other words, only combinations above and to the right of the red band lead to collapse. For velocities $v_{_{\rm O}}\ga 1.5\,{\rm km}\,{\rm s}^{-1}$, the convergence parameter must be $c\ga 0.6$, and for lower velocities it must be even greater.

The blue band on Fig. \ref{fig:c_crit} represents the same information for the more massive core, $M_{_{\rm CORE}}=0.060\,{\rm M}_{_\odot}$. With the higher mass, somewhat lower combinations of $v_{_{\rm O}}$ and $c$ are required for collapse, but the pattern is the same. At velocities $v_{_{\rm O}}\ga 1.0\,{\rm km}\,{\rm s}^{-1}$, the convergence parameter must be $c\ga 0.4$, and for lower velocities it must be even greater.

We have tested that these results are converged by repeating some of the critical simulations with four times as many SPH particles, i.e. ${\cal N}_{_{\rm SPHERE}}\sim 400000$. These simulations are marked with black dots on Fig. \ref{fig:c_crit}, and indicate that the simulations defining the red and blue bands are indeed well converged.

\begin{figure*}
\centering
\includegraphics[width=\textwidth]{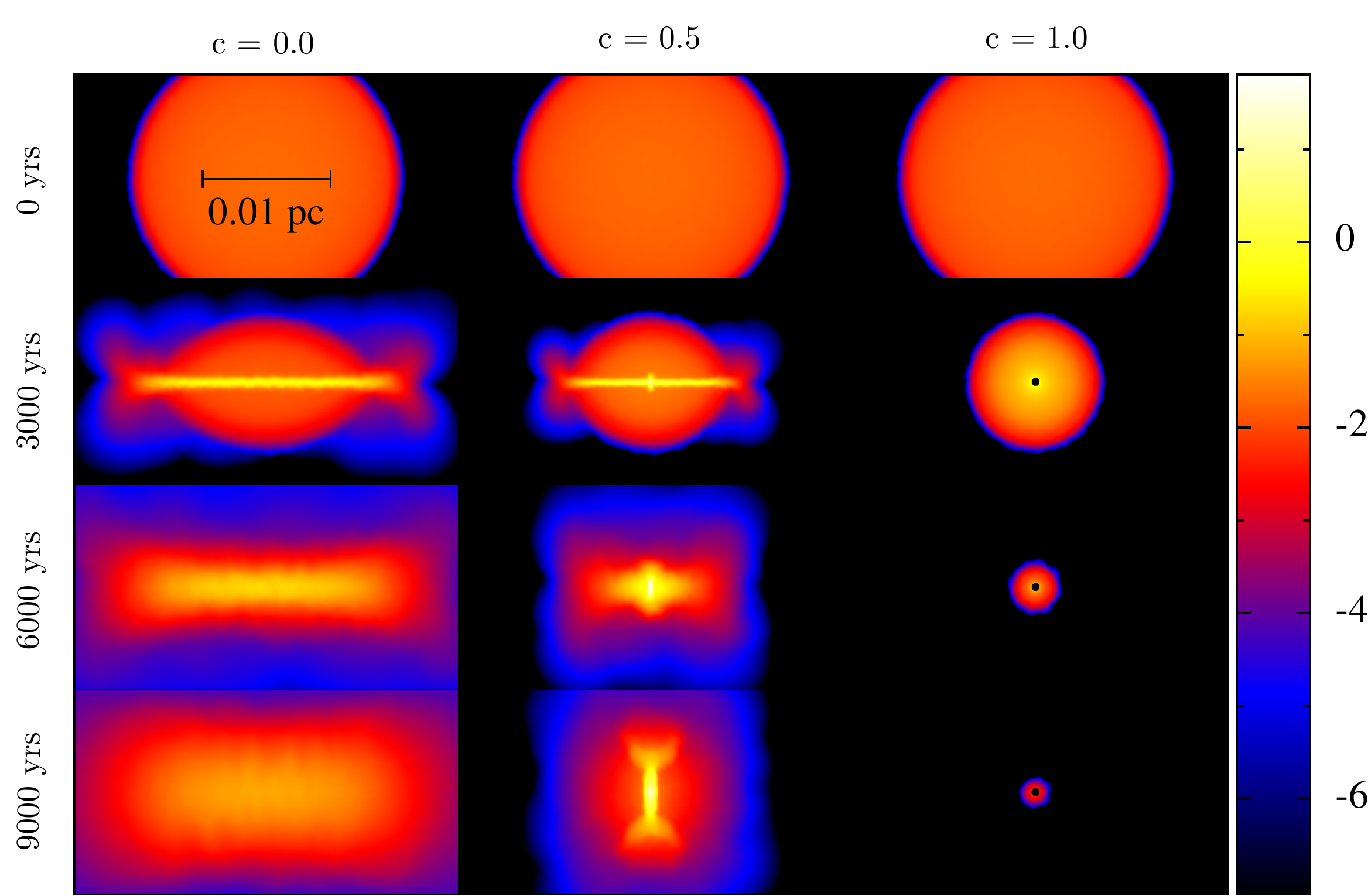}
\caption{False-colour column-density images of a core with mass $M=0.020\,{\rm M}_{_\odot}$, initial radius $R=0.010\,{\rm pc}$, and initial speed $v_{_{\rm O}}=2\,{\rm km}\,{\rm s}^{-1}$, as seen looking along the $y$-axis; the $z$-axis is vertical. In the left column $c=0$; in the central column $c=0.5$; and in the right column $c=1$. The frames on the top row show the initial conditions ($t=0$). The frames on the rows below show the core at $t=3000,\;6000\;{\rm and}\;9000\,{\rm yr}$ (reading from top to bottom). The black dot on the lower three frames of the right column ($c=1$) mark where a sink particle has formed, and is growing by accretion.}
\label{FIG:SideView}
\end{figure*}

\begin{figure*}
\centering
\includegraphics[width=\textwidth]{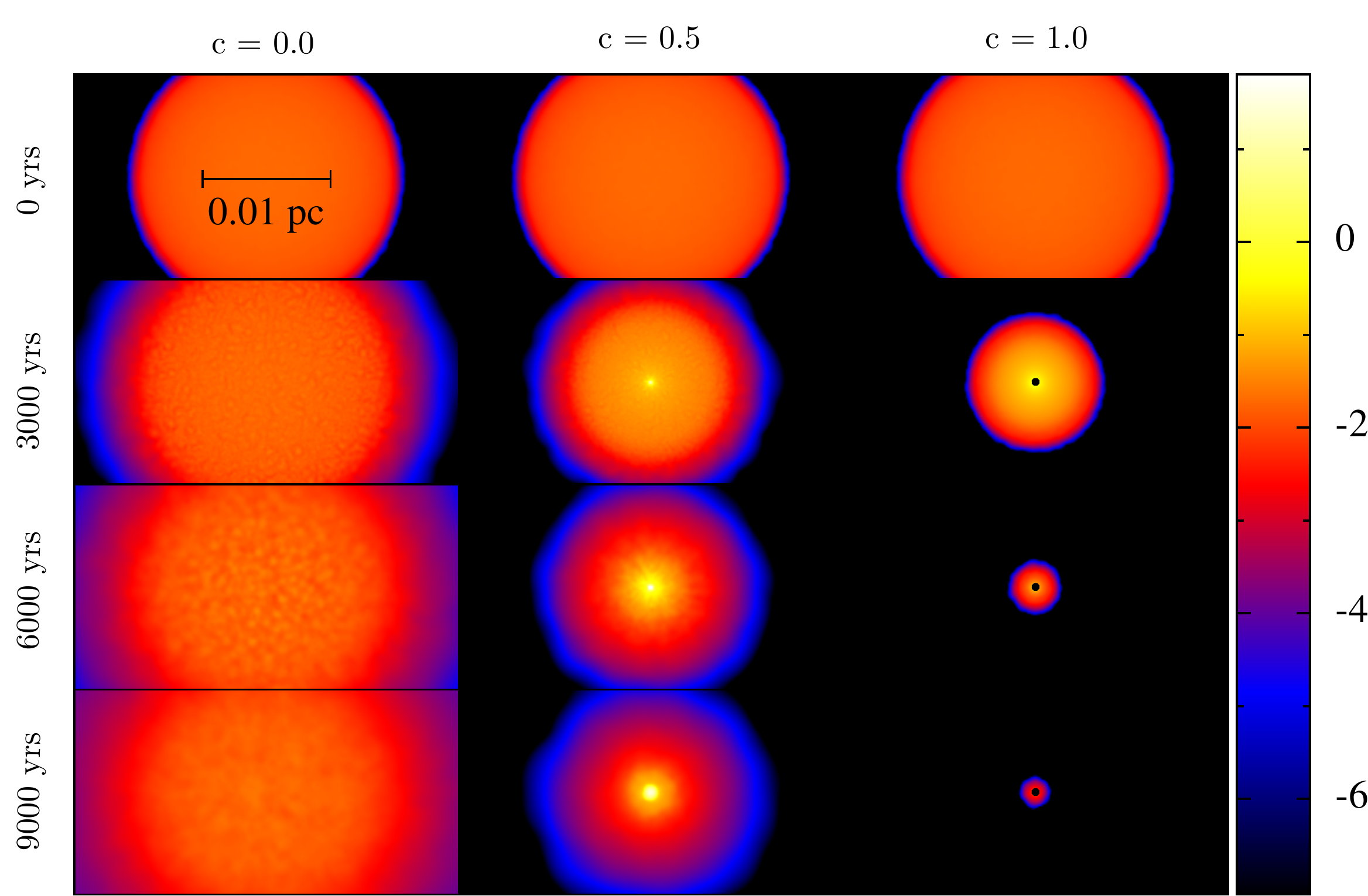}
\caption{As for Fig. \ref{FIG:SideView}, but looking along the $z$-axis.}
\label{FIG:HeadOnView}
\end{figure*}

Figs. \ref{FIG:SideView} and \ref{FIG:HeadOnView} present false-colour column-density images illustrating representative cases when the core mass is $M=0.020\,{\rm M}_{_\odot}$, its initial radius is $R=0.010\,{\rm pc}$, and the initial speed is $v_{_{\rm O}}=2\,{\rm km}\,{\rm s}^{-1}$. Fig. \ref{FIG:SideView} shows side views, i.e. projections on the $(x,z)$-plane, as seen looking along the $y$-axis; the $z$-axis is vertical. The left column represents the $c=0$ case, i.e. a head-on collision of two hemispheres; the central column represents the intermediate $c=0.5$ case; and the right column represents the $c=1$ case, i.e. purely radial inflow. From top to bottom the frames are at $t=0,\;3000,\;6000\;{\rm and}\;9000\,{\rm yr}$. Fig. \ref{FIG:HeadOnView} shows, in the same format, the corresponding views as seen looking along the $z$-axis. 

\begin{figure*}
\centering
\includegraphics[width=\textwidth]{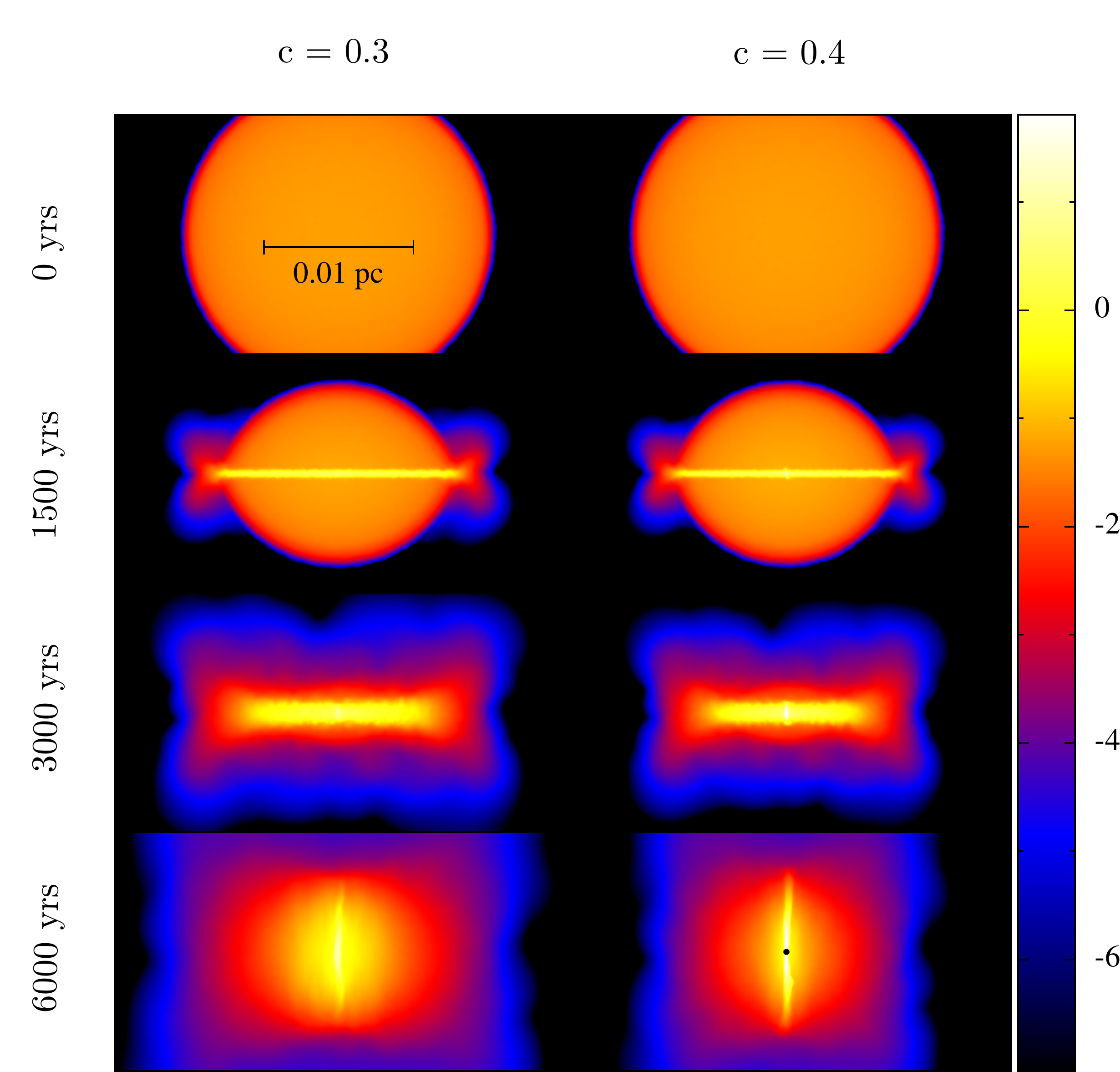}
\caption{False-colour column-density images from simulations performed with four times the standard number of SPH particles (i.e. ${\cal N}_{_{\rm SPHERE}}=400000$, as opposed to ${\cal N}_{_{\rm SPHERE}}=100000$), to demonstrate that the results are converged. All frames are as seen looking along the $y$-axis; the $z$-axis is vertical. In these cases the core has mass mass $M=0.060\,{\rm M}_{_\odot}$, initial radius $R=0.010\,{\rm pc}$, and initial speed $v_{_{\rm O}}=3\,{\rm km}\,{\rm s}^{-1}$. In the left column $c=0.3$, and a brown dwarf does not form. In the right column $c=0.4$, and a brown dwarf does form. The frames on the top row show the initial conditions ($t=0$). The frames on the rows below show the core at $t=1500,\;3000\;{\rm and}\;6000\,{\rm yr}$ (reading from top to bottom). The black dot on the lowest frame of the right column ($c=0.4$) marks where a sink particle has formed.}
\label{FIG:Convergence}
\end{figure*}

Fig. \ref{FIG:Convergence} presents false-colour column-density images illustrating representative cases when the core mass is $M=0.060\,{\rm M}_{_\odot}$, its initial radius is $0.010\,{\rm pc}$, and the initial speed is $v_{_{\rm O}}=3\,{\rm km}\,{\rm s}^{-1}$. These are simulations performed with ${\cal N}_{_{\rm SPHERE}}=400000$ SPH particles, i.e. four times as many as the standard simulations, to check convergence. The left column represents the $c=0.3$ case, which does not form a brown dwarf; the right column represents the $c=0.4$ case, which does form a brown dwarf. From top to bottom, the frames are at $t=0,\;1500,\;3000\;{\rm and}\;6000\,{\rm yr}$.

\subsection{Anti-parallel colliding flows, $c=0$}\label{SSEC:c=0}

When $c=0$, i.e. anti-parallel colliding flows, a brown dwarf is never formed, no matter how high the collision velocity, $v_{_{\rm O}}$. As shown in the left column of Fig. \ref{FIG:SideView}, a shock-compressed layer forms about the equatorial plane ($z\sim 0$), but this then disperses, at first by squeezing out sideways, and then -- once the inflow abates -- by dispersing in all directions.

The layer does not fragment, either during the period while it is still contained by the ram pressure of the inflowing gas ($t\la 5000\,{\rm yr}$), or after the inflow abates. This is because, during both periods, the minimum wavelength for fragmentation is larger than the extent of the layer, and the growth time for the fastest-growing wavelength is much longer than the duration of the collision. Specifically, the maximum surface-density of the layer is $\Sigma_{_{\rm MAX}}=M/\pi R^2$, and so the minimum fragmentation wavelength is
\begin{eqnarray}\nonumber
\lambda_{_{\rm MIN}}\;\sim\;\frac{a^2}{2G\Sigma_{_{\rm MAX}}}\!&\!\sim\!&\!\frac{\pi k_{_{\rm B}}TR^2}{2GM\bar{m}}\\
\!&\!\sim\!&\!0.2\,{\rm pc}\left(\!\frac{M}{0.02\,{\rm M}_{_\odot}}\!\right)^{-1}\left(\!\frac{T}{10\,{\rm K}}\!\right),
\end{eqnarray}
which is larger than the lateral extent of the collision interface ($\sim\!0.04\,{\rm pc}$). Moreover, the fastest growing fragmentation wavelength, $\lambda\simeq 2\lambda_{_{\rm MIN}}$, has a growth time of
\begin{eqnarray}\nonumber
t_{_{\rm FASTEST}}&\sim&\frac{a}{2G\Sigma_{_{\rm MAX}}}\\
&\sim&1\,{\rm Myr}\left(\!\frac{M}{0.02\,{\rm M}_{_\odot}}\!\right)^{-1}\left(\!\frac{T}{10\,{\rm K}}\!\right)^{1/2}\,,
\end{eqnarray}
which is much longer than the duration of the collision ($\la 0.01\,{\rm Myr}$). This is why there is no significant tendency for material to condense towards the centre of the layer, and hence no indication of a central density peak developing in the bottom left frame of Fig. \ref{FIG:HeadOnView} ($c=0, t=9000\,{\rm yr}$). In making the above estimates of $\lambda_{_{\rm MIN}}$ and $t_{_{\rm FASTEST}}$, we have assumed that the shock-compressed gas in the layer cools down to $\sim\!10\,{\rm K}$ very quickly, which is usually a valid assumption (Whitworth 2015, submitted), but the conclusions would be even stronger if the gas did not cool so quickly, since there would then be an even greater over-pressure driving dispersal of the shock-compressed gas in directions not contained by the ram pressure of the inflow.

\subsection{Intermediate cases, $0<c<1$}

In the intermediate cases, some of the inflow velocity is invested in inward motions that are not parallel (or anti-parallel) to the $z$-axis, and therefore deliver equatorial contraction from the outset. Consequently the flow converges somewhat more slowly on the $z\!=\!0$ plane, but the extent of the collision interface also shrinks, as shown in the central columns of Figs. \ref{FIG:SideView} and \ref{FIG:HeadOnView}. Consequently there is a trade-off between the inward equatorial motions that are trying to form a gravitationally unstable condensation at the centre, and the termination of the inflows parallel and anti-parallel to the $z$-axis, which relieves the high ram-pressure containing the layer, and allows the layer to expand in the polar directions.

In the case illustrated in the central columns of Figs. \ref{FIG:SideView} and \ref{FIG:HeadOnView} ($M=0.020\,{\rm M}_{_\odot}$, $R=0.010\,{\rm pc}$, $v_{_{\rm O}}=2\,{\rm km}\,{\rm s}^{-1}$, $c=0.5$), a condensation starts to form, but does not collapse to form a brown dwarf, because, once the inflows terminate, the condensation stops contracting, expands along the $z$-axis and disperses.

The cases illustrated in Fig. \ref{FIG:Convergence} ($M=0.060\,{\rm M}_{_\odot}$, $R=0.010\,{\rm pc}$, $v_{_{\rm O}}=3\,{\rm km}\,{\rm s}^{-1}$) straddle the switch between failed brown-dwarf formation (left column, $c=0.3$) and successful brown-dwarf formation (right column, $c=0.4$). The extra equatorial convergence in the $c=0.4$ case is able to form a brown dwarf (i.e. a gravitationally bound object) before the inflow terminates (and with it the containing ram pressure); matter still then flows away along the polar directions, limiting the mass of the brown dwarf, but it is too late to stop the brown dwarf from forming. 

\subsection{Purely radial inflow, $c=1$}

When $c=1$, i.e. a purely radial flow, a brown dwarfs forms if $v_{_{\rm O}}\gtrsim 1.0\,\mathrm{km\,s^{-1}}$, when $M_\textsc{core}=0.020\,\mathrm{M_\odot}$; and if $v_{_{\rm O}}\gtrsim 0.5\,\mathrm{km\,s^{-1}}$, when $M_\textsc{core}=0.060\,\mathrm{M_\odot}$. The case for $M=0.020\,{\rm M}_{_\odot}$, $R=0.010\,{\rm pc}$, $v_{_{\rm O}}=2\,{\rm km}\,{\rm s}^{-1}$ and $c=1$ is illustrated in the right columns on Figs. \ref{FIG:SideView} and \ref{FIG:HeadOnView}. We see that, in this case, a brown dwarf has formed by $t=3000\,{\rm yr}$. Since, apart from fluctuations due to particle noise, this simulation is spherically symmetric, the right columns of Figs.  \ref{FIG:SideView} and \ref{FIG:HeadOnView} are indistinguishable.

When $c=1$, the spherical symmetry also means that there is no preferred direction of escape for the over-pressured gas in the central regions of the core. Therefore the only way a core with $c=1$ can avoid forming a brown dwarf is if (i) the contraction is brought to a halt, because the work done against compression equals the initial inward kinetic energy plus the gravitational potential energy released, and (ii) the core is then unbound and disperses.

\begin{figure}
\centering
\includegraphics[width=\columnwidth]{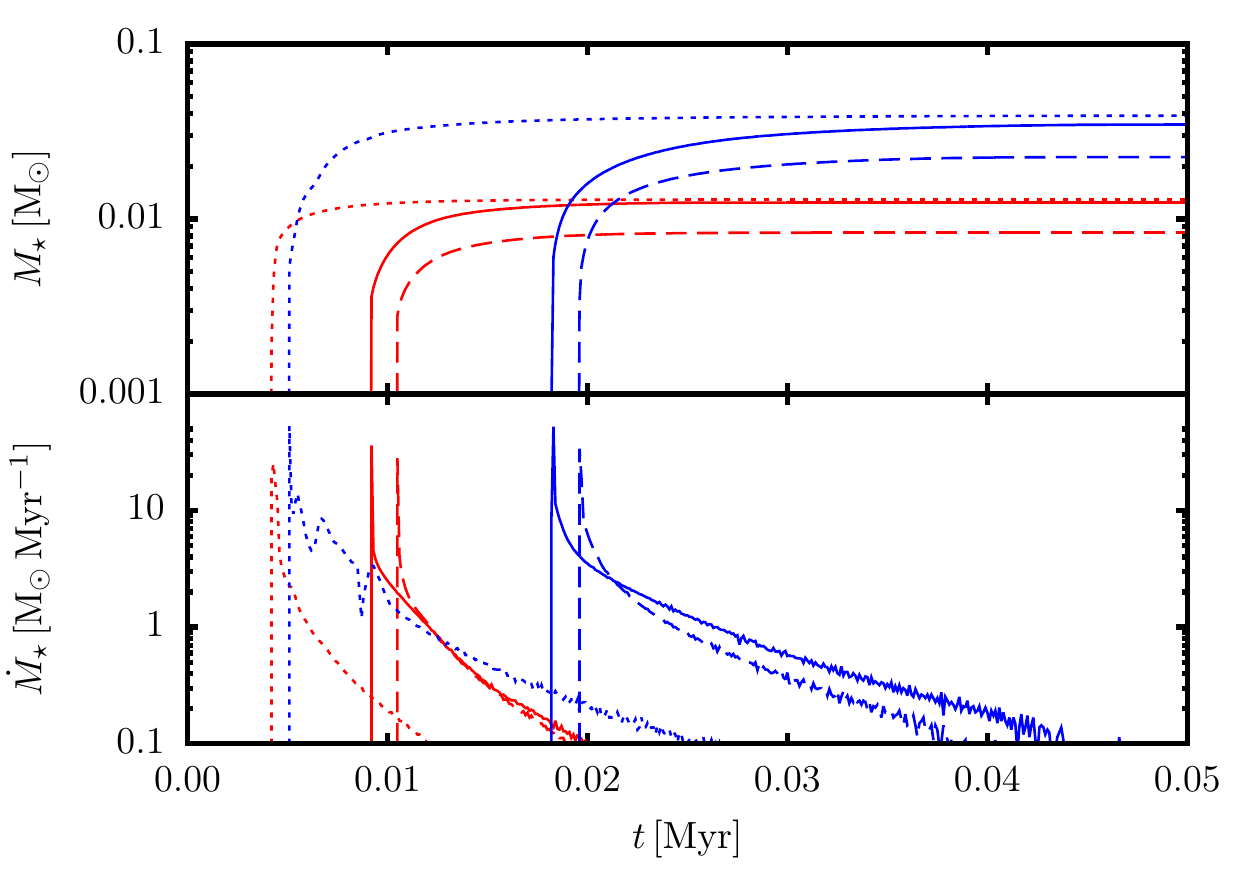}
\caption{(Top) Sink mass as a function of time for the high-resolution simulations that form sinks. The red lines are for cores with $M_{_{\rm CORE}}=0.020\,{\rm M}_{_\odot}$, and the blue lines are for cores with $M_{_{\rm CORE}}=0.060\,{\rm M}_{_\odot}$: red solid line, $(v_{_{\rm O}},c)=(1.0\,{\rm km}\,{\rm s}^{-1},0.9)$; red dashed line, $(1.5\,{\rm km}\,{\rm s}^{-1},0.6)$; red dotted line, $(3.0\,{\rm km}\,{\rm s}^{-1},0.6)$; blue solid line, $(0.5\,{\rm km}\,{\rm s}^{-1},0.7)$; blue dashed line, $(1.0\,{\rm km}\,{\rm s}^{-1},0.4)$; blue dotted line, $(3.0\,{\rm km}\,{\rm s}^{-1},0.4)$. (Bottom) As before, but with sink accretion rates.}
\label{FIG:Mdot}
\end{figure}

\subsection{Final brown dwarf masses}

Fig. \ref{FIG:Mdot} shows the sink masses as a function of time for the high-resolution simulations that form sinks (i.e. the ones represented by black dots on the upper boundaries of the coloured bands on Fig. \ref{fig:c_crit}. The red lines are for cores with $M_{_{\rm CORE}}=0.020\,{\rm M}_{_\odot}$, and the blue lines are for cores with $M_{_{\rm CORE}}=0.060\,{\rm M}_{_\odot}$. We note that the corresponding minimum resolvable masses are $0.00002\,{\rm M}_{_\odot}$ and $0.00006\,{\rm M}_{_\odot}$, so the formation, growth and saturation of the brown-dwarf masses is well resolved.

\subsection{Other configurations}

We have repeated these simulations, using cores with the same initial density fields (i.e. masses of $0.020\,{\rm M}_{_\odot}$ and $0.060\,{\rm M}_{_\odot}$, initial radii of $0.010\,{\rm pc}$, uniform initial densities), but initial velocity fields that have no $x$-component. Specifically, the initial velocity field is now given by
\begin{equation}
 \boldsymbol{v}=-\,v_{_{\rm O}}\,\frac{c\,\hat{\boldsymbol{r}}'+(1-c)\,\hat{\boldsymbol{z}}}{|c\,\hat{\boldsymbol{r}}'+(1-c)\,\hat{\boldsymbol{z}}|}\,,
\end{equation}
where, $\hat{\boldsymbol{r}}'=(0,y,z)/\sqrt{y^2+z^2}$. When $c=0$, the system comprises two hemispheres colliding head-on, just as with $c=0$ in Section \ref{SSEC:c=0}. When $c=1$, the velocity field is cylindrically symmetric about the $x$-axis, and independent of $x$.

We have performed simulations with this initial velocity field, using the same combinations of $v_{_{\rm O}}$ and $c$ as are listed at the start of Section \ref{SSSEC:Overview}; in no case does a brown dwarf form. Material can always squirt out in the $\pm x$-directions, and so growth of the central density is suppressed.

\subsection{Interim summary}

The inference that we draw from these simulations is that the formation of an isolated brown dwarf by turbulent fragmentation requires flows that converge from all directions. It is not obvious that such flows will arise very frequently in nature. There must therefore be some doubt as to whether many isolated brown dwarfs form by turbulent fragmentation. In the next section we revisit the derivation of the mass of Oph-B11, and conclude that, although it is likely to be below the hydrogen-burning limit, this is not certain. In Section \ref{SEC:ALTS} we consider alternative explanations for Oph-B11.

\section{The mass of Oph-B11}\label{SEC:MASS}

The flux density ({\it aka} monochromatic flux), $S_\nu$, from an optically thin, isothermal cloud is a function of five parameters:
\begin{equation}
  S_\nu(M,D,\kappa_{\nu_0},T,\beta)=\frac{M\,\kappa_{\nu_0}\,B_\nu(T)}{D^2}\left(\frac{\nu}{\nu_0}\right)^\beta\,.
  \label{eqn:greybody}
\end{equation}
Here $M$ is the total mass of the cloud, $D$ is the distance to the cloud, $T$ is the temperature, $\kappa_{\nu_0}$ is the monochromatic mass opacity coefficient at frequency $\nu_0$ (i.e. the total cross section presented by unit mass of gas and dust to photons of frequency $\nu_{_{\rm O}}$), $\beta$ is the emissivity index (i.e. $\beta=d\ln[\kappa_\nu]/d\nu$), and $B_\nu(T)$ is the Planck function. Like \citet{AWG12}, we fit Eqn. (\ref{eqn:greybody}) to the flux density measurements for Oph-B11 in order to estimate its mass. Unlike \citet{AWG12}, we allow all five parameters to vary -- within limits which we discuss below.

We use a Bayesian analysis to infer the posterior Probability Density Function (PDF) for the mass of Oph-B11. This inference is made by calculating the posterior PDF for all the parameters in Eqn. (\ref{eqn:greybody}) and marginalising out all of them except $M$. The posterior PDF for the parameters $\boldsymbol{\theta}=(M,D,\kappa_{\nu_0},T,\beta)$, given data $\boldsymbol{S}=(S_1,S_2,\dots S_n)$, is formally defined by
\begin{equation}
P(\boldsymbol{\theta}|\boldsymbol{S})=\frac{P(\boldsymbol{S}|\boldsymbol{\theta})\,P(\boldsymbol{\theta})}{P(\boldsymbol{S})}\,.
\label{eqn:bayes}
\end{equation}
Here, $P(\boldsymbol{S}|\boldsymbol{\theta})$ is the likelihood of measuring $\boldsymbol{S}$, given $\boldsymbol{\theta}$;
$P(\boldsymbol{\theta})$ is the prior PDF of $\boldsymbol{\theta}$; and $P(\boldsymbol{S})$ is the normalisation constant that ensures $\int\!P(\boldsymbol{\theta}|\boldsymbol{S})\,\mathrm{d}\boldsymbol{\theta}\equiv 1$\,.

We note that sophisticated Bayesian methods for extracting the masses of cores already exist \citep[e.g.][]{KSS12,VPN13}. However (i) we do not have access to the reduced data and (ii) in any case we are more interested in how the \emph{a priori} uncertainties on $D$, $\kappa_0$, $T$ and $\beta$ combine to affect the \emph{a posteriori} uncertainty on $M$.

\subsection{Data}

The flux density data used by \citet{AWG12} to infer the mass of Oph-B11 comprise two detections and four non-detections. There are flux densities at $3.2\,\mathrm{mm}$ and $850\,\mathrm{\upmu m}$, measured -- respectively -- with the Institut de Radioastronomie Millim\'{e}trique (IRAM) Plateau de Bure Interferometer (PdBI), and with the Submillimetre Common-User Bolometer Array (SCUBA) instrument of the James Clerk Maxwell Telescope (JCMT) telescope. In addition there are four $3\sigma$ upper limits obtained with the Spectral and Photometric Imaging REceiver (SPIRE) and Photodetector Array Camera and Spectrometer (PACS) on the Herschel Space Observatory at $250,\,160,\,110\;{\rm and}\;70\,\mathrm{\upmu m}$. We define a six-element vector $\boldsymbol{S}$ in which we store these data. There is a second six-element vector $\boldsymbol{\sigma}$ in which we store the corresponding uncertainties (a combination of confusion noise and uncertainty on the background cirrus subtraction). Values of $S_i$ and $\sigma_i$ are given in Table \ref{TAB:ObsData}.

\begin{table}
\begin{tabular}{cccccc}\hline
 $i$ & $\lambda_i\,[\mathrm{\upmu m}]$ & $S_i\,[\mathrm{mJy}]$ & $\sigma_i\,[\mathrm{mJy}]$ & Telescope & Instrument \\\hline
 1 & 3200 & 0.4 & 0.1 & IRAM & PdBI \\
 2 & 850 & 39 & 5 & JCMT & SCUBA \\
 3 & 250 & $\leq390$ & 130 & Herschel & SPIRE \\
 4 & 160 & $\leq170$ & 57 & Herschel & PACS \\
 5 & 110 & $\leq36$ & 12 & Herschel & PACS \\
 6 & 70 & $\leq83$ & 28 & Herschel & PACS \\\hline
\end{tabular}
\caption{Oph-B11 flux densities at various wavelengths. Rows 1 and 2 are detections. Rows 3 to 6 are $3\sigma$ non-detections, i.e. upper limits.}
\label{TAB:ObsData}
\end{table}

\subsection{Likelihood function}\label{SSEC:Likelihood}

The likelihood of measuring flux densities $\boldsymbol{S}$, given parameters $\boldsymbol{\theta}$, is:
\begin{equation}
  P(\boldsymbol{S}|\boldsymbol{\theta}) = \prod_{i=1}^{i=6}\left\{P(S_i|\boldsymbol{\theta})\right\}\,.
\end{equation}

Flux densities $S_1$ and $S_2$ are detections. If we assume that they have Gaussian uncertainties, the likelihood of measuring $S_i$ is given by,
\begin{equation}
  P(S_i|\boldsymbol{\theta})=\frac{1}{(2\pi)^{1/2}\sigma_i}\,\exp\left(-\frac{[S_i-S_\nu(\boldsymbol{\theta})]^2}{2\,\sigma_i^2}\right)\,,
\end{equation}
where $S_\nu(\boldsymbol{\theta})$ is given by Eqn. (\ref{eqn:greybody})\,.

Flux densities $S_3$, $S_4$, $S_5$ and $S_6$ are non-detections, i.e. upper limits. If we again assume Gaussian uncertainties, the likelihood of measuring upper limit $S_i$ is given by
\begin{equation}
  P(S_i|\boldsymbol{\theta})=\frac{1}{2}+\frac{1}{2}\,\mathrm{erf}\left(\frac{S_i-S_\nu(\boldsymbol{\theta})}{\sqrt{2}\,\sigma_i}\right)\,,
\end{equation}
where $\mathrm{erf}(x)$ is the error function,
\begin{equation}
  \mathrm{erf}(x)\equiv\frac{1}{\sqrt{\uppi}}\int_0^x \exp(-x^2)\,\mathrm{d}x\,.
\end{equation}
Here, $P(S_i|\boldsymbol{\theta})$ is the fraction of the distribution of $S_\nu(\boldsymbol{\theta})\pm\sigma_i$ below $S_i$.

\subsection{Prior probability density function}\label{SSEC:Prior}

\begin{table}
\centering{
\setlength{\extrarowheight}{3.0pt}
\begin{tabular}{ccc}\hline
 $\theta$ & $\mu_\theta$ & $\sigma_\theta$ \\\hline
 $M$ & -- & -- \\
 $D$ & $140\,\mathrm{pc}$ & $21\,\mathrm{pc}$ \\
 $T$ & $9\,\mathrm{K}$ & $1\,\mathrm{K}$ \\
 $\kappa_{_{230\,\mathrm{GHz}}}$ & $5\!\times\! 10^{-3}\,{\rm cm}^2\,{\rm g}^{-1}$ & $\divideontimes2$ \\
 $\beta$ & $1.8$ & $0.2$ \\\hline
\end{tabular}}
\caption{Means and standard deviations for the parameter prior PDFs. Column 1 gives the parameter symbol, Column 2 the mean, and Column 3 the standard deviation.}
\label{TAB:Priors}
\end{table}

The prior PDF describes our \emph{a priori} preconceptions regarding credible values for the parameters $\boldsymbol{\theta}$. If the components $\theta_i$ are uncorrelated, the prior PDF is given by
\begin{equation}
P(\boldsymbol{\theta}) = \prod_{j=1}^{5}\left\{P(\theta_j)\right\}\,.
\end{equation}

\subsubsection{Prior PDF for the mass, $M$}

We consider three prior PDFs for $M$.

The first two are \emph{uninformative}, i.e. they make no \emph{a priori} assumptions concerning the mass of Oph-B11. The first is the {\it uniform} prior PDF, which postulates that the probability of a value between $M$ and $M+\mathrm{d}M$ is proportional to $dM$,
\begin{equation}
P_\textsc{uni}(M)\propto M^{0}\,.
\end{equation}
The second is the {\it logarithmic} prior PDF, which postulates that the probability of a value between $\log [M]$ and $\log [M]+\mathrm{d}\log [M]$ is proportional to $d\log [M]$,
\begin{equation}
P_\textsc{log}(M)\propto M^{-1}\,.
\end{equation}

The third prior PDF is \emph{informative}, and based on the observed core mass function (CMF). Observations suggest that the CMF is very similar in shape to the Initial Mass Function \citep[IMF; e.g.][]{KAM10}. The third prior PDF is therefore the {\it IMF} prior PDF, and postulates that the mass of Oph-B11 is drawn from the CMF, and can therefore be approximated by the slope of the low-mass tail of the \citet{K01} IMF,
\begin{equation}
P_\textsc{imf}(M)\propto M^{-0.3}\,.
\end{equation}

We note that if all other parameters are fixed, i.e. their prior PDFs are delta functions, then the posterior PDF of $M$ is dominated by the likelihood function, and the choice of prior PDF for $M$ is unimportant.

\subsubsection{Prior PDFs for other parameters}

We assume that each of the remaining four parameters, ($D,T,\kappa_{\nu_0}\text{ and }\beta$), subscribes to a Gaussian prior PDF,
\begin{eqnarray}
P(\theta)&=&\frac{1}{(2\pi)^{1/2}\sigma_\theta}\,\exp\left(-\frac{(\theta-\mu_\theta)^2}{2\,\sigma_\theta^2}\right)\,,
\end{eqnarray}
where the values of $\mu_\theta$ and $\sigma_\theta$ are given in Table \ref{TAB:Priors}. Following the assumptions of \citet{AWG12}, we use $D=140\pm21\,\mathrm{pc}$ \citep{M08} and $T=9\pm1\,\mathrm{K}$ \citep[e.g.][]{SWW-T08,LWL99}. We adopt the commonly used mass opacity coefficient, $\kappa_{230\,\mathrm{GHz}}=5.0\times10^{-3}\,\mathrm{cm^2\,g^{-1}}$ \citep{AWM96,MAN98}, which is uncertain by a factor of order two \citep[with lower values for naked dust grains and higher values for coagulated grains;][] {OH94}. Finally, we assume $\beta=1.8\pm0.2$, which corresponds to the value calculated by \citet{OH94}, $\beta\simeq 1.8$. The uncertainty accommodates the frequently used, slightly higher value $\beta=2.0$ \citep[e.g.][]{NW-T07,SNW-T08,AWG12}, as well as the lower limit $\beta\approx1.5$ estimated from the Planck all-sky survey \citep[e.g.][]{PLANCK11}.

\subsection{Posterior probability density function}\label{SSEC:Posterior}

We use a Markov-chain Monte Carlo method to sample the posterior PDF of $\boldsymbol{\theta}$. The chain begins at an arbitrary position $\boldsymbol{\theta}_k$. A candidate new position is then generated with 
\begin{equation}
 \boldsymbol{\theta}_{k+1}=\boldsymbol{\theta}_k\;+\;
 \begin{pmatrix}
  \Delta\theta_1, \\
  \Delta\theta_2, \\
  \vdots\\
  \Delta\theta_m
 \end{pmatrix}
 (\mathcal{G}_1,\mathcal{G}_2,\cdots\mathcal{G}_m)\,,
\label{EQN:Step}
\end{equation}
where the $\Delta\theta_j$ are pre-set random-walk step-lengths and the $\mathcal{G}_j$ are random numbers drawn from the Gaussian distribution with zero mean and unit standard deviation. This candidate new position is adopted if
\begin{equation}
 \frac{P(\boldsymbol{D}|\boldsymbol{\theta}_{k+1})\,P(\boldsymbol{\theta}_{k+1})}{P(\boldsymbol{D}|\boldsymbol{\theta}_{k})\,P(\boldsymbol{\theta}_{k})}>\mathcal{R}\,,
\label{eqn:chain_chriterion}
\end{equation}
where $\mathcal{R}$ is a random number from the uniform distribution between zero and one. If Eqn. (\ref{eqn:chain_chriterion}) is not satisfied, $\boldsymbol{\theta}_{k+1}$ is discarded and a new candidate position is generated using Eqn. (\ref{EQN:Step}). This process is repeated iteratively to create a long chain of positions in $\boldsymbol{\theta}$-space.

An optimal choice for $\Delta\theta_j$ is the posterior standard deviation of $\theta_j$. This information is not available at the outset, so we start off with estimates based on the prior distributions, i.e. $\Delta\theta_j=\sigma_{\theta_j}$. Next, we perform a \emph{burn-in} by computing a chain of $10^4$ points, and then revising the set of $\Delta\theta_j$ values to the standard deviations of the resulting $10^4$ $\theta_j$ values. Finally, the points from the burn-in are discarded, and a chain of $10^6$ points is computed using these $\Delta\theta_j$. The resulting $10^6$ points, $\boldsymbol{\theta}_k$, approximate an ensemble of points drawn randomly from the posterior PDF, $P(\boldsymbol{\theta}|\boldsymbol{D})$.

\subsection{Results}

\subsubsection{Mass distribution}

\begin{figure}
\centering
\includegraphics[width=\columnwidth]{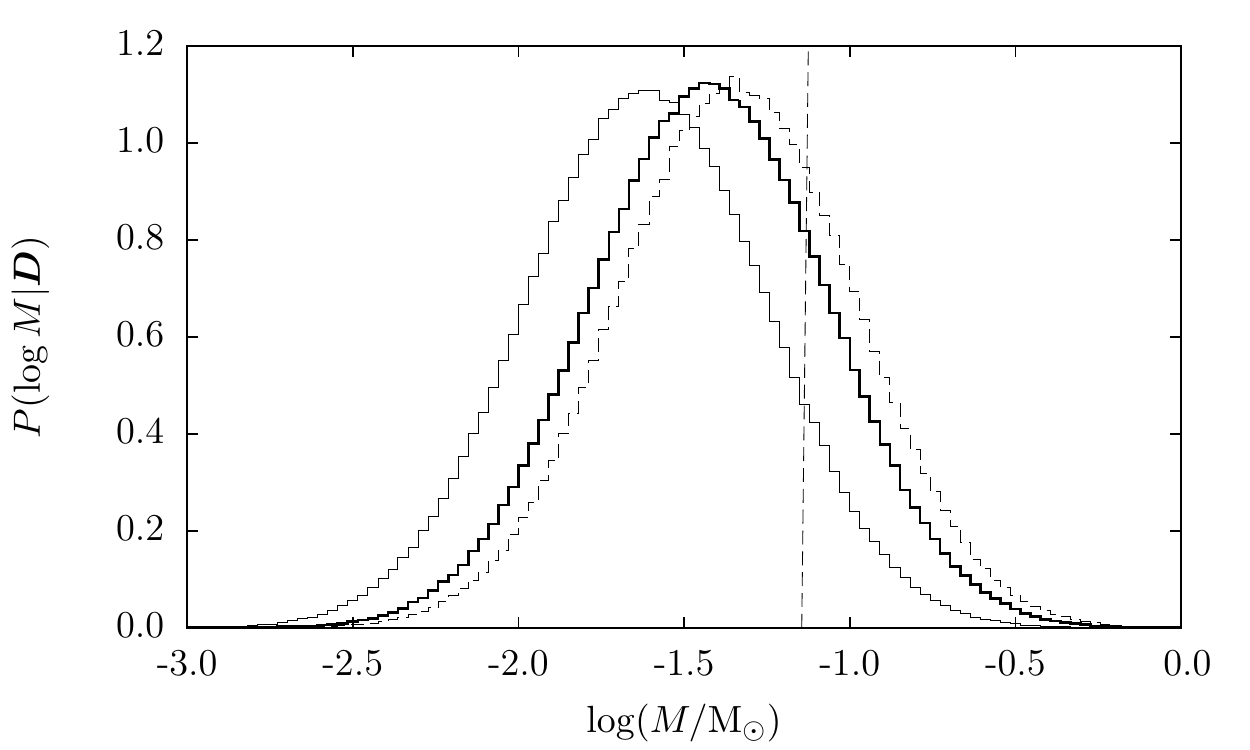}
\caption{The posterior PDF of $\log_{_{10}}[M]$. The solid histogram (in the centre) shows the posterior PDF when the $P_\textsc{imf}(M)$ prior PDF is used. The dashed histogram (to the right) shows the posterior PDF when the $P_\textsc{uni}(M)$ prior PDF is used. The dotted histogram (to the left) shows the posterior PDF when the $P_\textsc{log}(M)$ prior PDF is used. The vertical dotted grey line shows the hydrogen burning limit at $\sim\!0.075\,\mathrm{M_\odot}$.}
\label{fig:1d_hist}
\end{figure}

\begin{figure}
\centering
\includegraphics[width=\columnwidth]{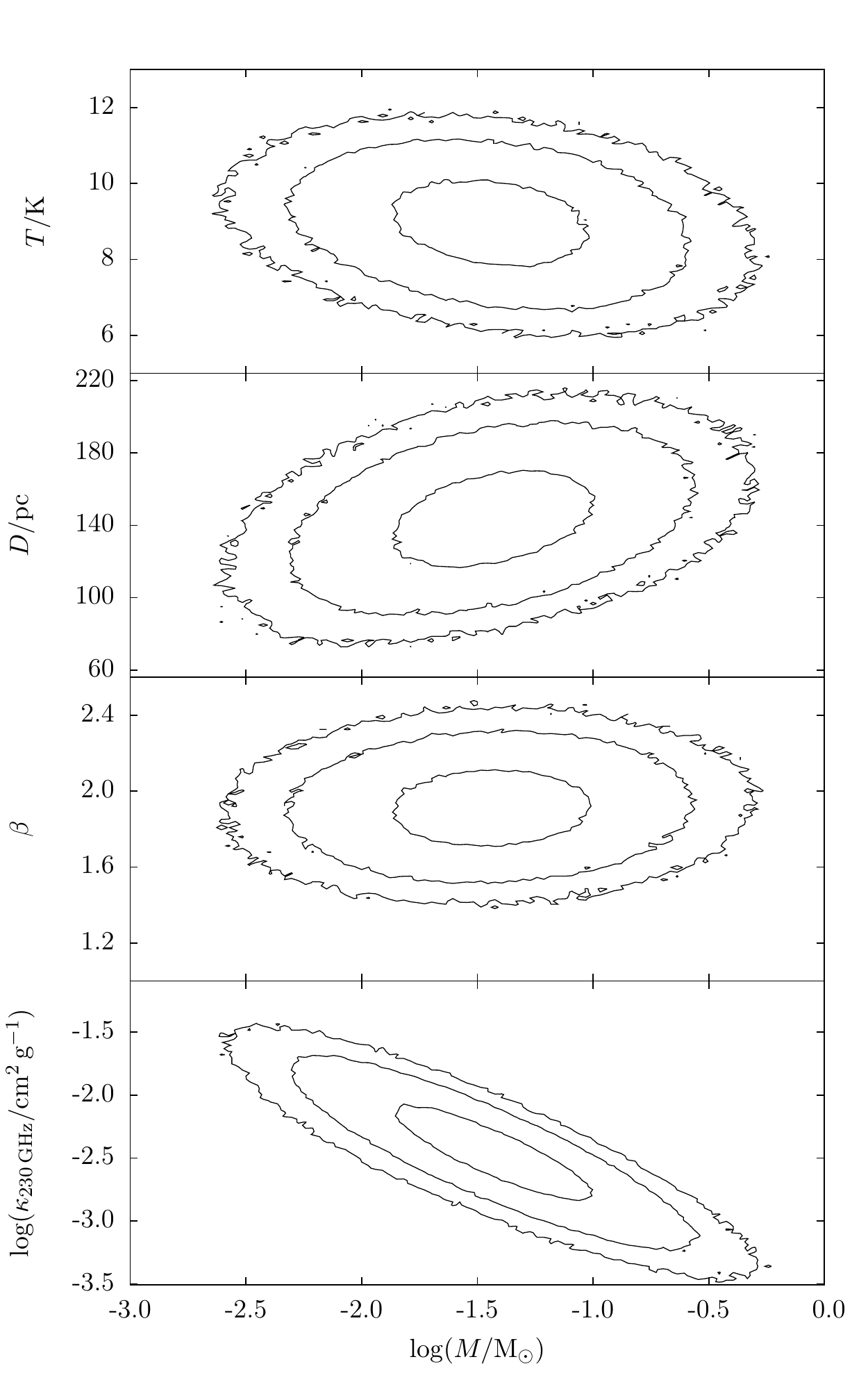}
\caption{From top to bottom: the posterior PDFs of $\log_{_{10}}[M]$ and $T$; $\log_{_{10}}[M]$ and $D$; $\log_{_{10}}[M]$ and $\beta$; and $\log_{_{10}}[M]$ and $\log_{_{10}}[\kappa_{_{230\,{\rm GHz}}}]$. These posterior PDFs are calculated using the $P_\textsc{imf}(M)$ prior PDF. The contours encircle $50\%$, $95\%$ and $99.5\%$ of the distribution (from the inner-most to the outer-most contour).}
\label{fig:2d_hist}
\end{figure}

\begin{figure}
\centering
\includegraphics[width=\columnwidth]{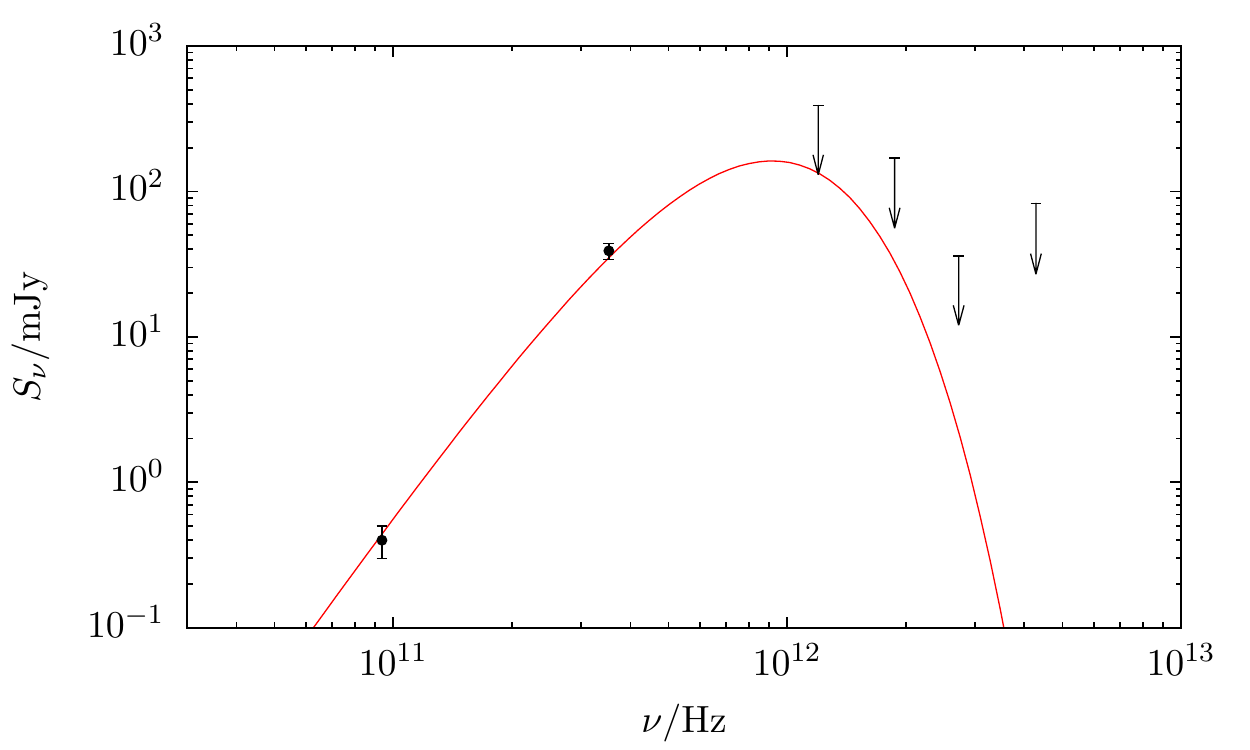}
\caption{Oph-B11 flux densities with a modified blackbody fit (see Eqn. \ref{eqn:greybody}). For the fit, $T=9\,\mathrm{K}$, $\beta=1.9$ and $M\,\kappa_{_{230\,{\rm GHz}}}/D^2=1.3\times10^{-12}$. The points with error bars are detections. The points with downwards arrows are $3\sigma$ non-detections.}
\label{fig:greybody}
\end{figure}

\begin{table}
\centering{
\setlength{\extrarowheight}{3.0pt}
\begin{tabular}{ccc}\hline
 Prior & $M/0.01\,{\rm M}_{_\odot}$ & $P_\textsc{bd}$ \\\hline
 $P_\textsc{imf}(M)$ & $3.8^{+4.8}_{-2.1}$ & 0.73 \\
 $P_\textsc{uni}(M)$ & $4.5^{+5.8}_{-2.5}$ & 0.66 \\
 $P_\textsc{log}(M)$ & $2.4^{+3.1}_{-1.4}$ & 0.86 \\\hline
\end{tabular}}
\caption{Estimated mass of Oph-B11. Column 1 gives the prior PDF used for the mass. Column 2 give the geometric mean mass and its standard deviation. Column 3 gives the probability of Oph-B11 having a brown-dwarf mass, i.e. below the hydrogen-burning limit at $\sim 0.075\,\mathrm{M_{_\odot}}$.}
  \label{tbl:mass_results}
\end{table}

Fig. \ref{fig:1d_hist} shows the three mass distributions obtained with the three different prior PDFs for $M$. The geometric means plus their standard deviations, and the fraction of cores, $P_{_{\rm BD}}$, that fall below the hydrogen-burning limit, are summarised in Table \ref{tbl:mass_results}\,. Using the $P_\textsc{imf}(M)$ prior, we find that there is roughly a 73\% probability that the mass of Oph-B11 is below $\sim\!0.075\,\mathrm{M_\odot}$. With the $P_\textsc{uni}(M)$ prior, which is biased towards higher-mass objects, there is a 66\% probability that the mass is below $\sim\!0.075\,\mathrm{M_\odot}$. With the $P_\textsc{log}(M)$ prior, which is biased towards lower-mass objects, there is a 86\% probability that the mass is below $\sim\!0.075\,\mathrm{M_\odot}$. This contrasts with the analysis of \citet{AWG12}, who conclude that the mass of Oph-B11 cannot be much higher than $\sim\!0.030\,\mathrm{M_\odot}$.

\subsubsection{Relationship between mass and other parameters}

Fig. \ref{fig:2d_hist} shows the posterior PDFs of $(\log_{_{10}}[M],T)$, $(\log_{_{10}}[M],D)$, $(\log_{_{10}}[M],\beta)$ and $(\log_{_{10}}[M],\log_{_{10}}[\kappa_{_{230\,{\rm GHz}}}])$, obtained with the $P_\textsc{imf}(M)$ prior PDF. Unsurprisingly, we find that the estimated mass, $M$, is correlated with the adopted distance, $D$, and anti-correlated with both the adopted temperature, $T$, and the adopted dust mass opacity coefficient, $\kappa_{_{230\,{\rm GHz}}}$. There is no strong relationship between the estimated mass, $M$, and the adopted dust emissivity index, $\beta$.

Fig. \ref{fig:greybody} shows the observed Spectral Energy Distribution of Oph-B11 (two points and four upper limits). The curve is the best modified-blackbody fit, and is obtained with $T=9\,\mathrm{K}$, $\beta=1.9\,$ and $\,M\kappa_{_{230\,{\rm GHz}}}/D^2=1.3\times10^{-12}$.

\subsection{Discussion}

\citet{AWG12} report that Oph-B11 has a mass between $0.020\,{\rm M}_{_\odot}$ and $0.030\,\mathrm{M_\odot}$. Our analysis is compatible with a mass in this range. However, when account is taken of the uncertainties on the physical parameters, there is only a $\sim\! 20\%$ probability that the mass falls in this range. Furthermore, it is more likely that the mass falls above this range than below it. The expectation value only falls in the range $0.020\,{\rm M}_{_\odot}$ to $0.030\,\mathrm{M_\odot}$ if we adopt the $P_\textsc{log}(M)$ prior PDF, i.e. we assume that the prior probability of finding a prestellar core with log-mass between $\log[M]$ and $\log[M]+\mathrm{d}\log[M]$ is proportional to $d\log[M]$. We know from observations of prestellar cores that this is unlikely, i.e. the CMF appears to be similar in shape to the IMF and not log-uniform. If we factor this information into our analysis of Oph-B11, we find that the mass has an expectation value of $\sim\!0.04\,\mathrm{M_\odot}$, and that there is a $\sim\!27\%$ probability that the mass is above the hydrogen-burning limit.

\subsubsection{Caveats}

We note the following three limitations of our analysis

First, we assume that the uncertainties on the data and the prior parameters have Gaussian distributions. This is the distribution with the maximum entropy -- and hence minimum presumed information -- for any quantity with a mean and a variance. 

Second, we do not include any correlations between the prior PDFs. Such correlations are difficult to establish. For example, \citet{KSS12} report a weak positive correlation between $T$ and $\beta$ on the basis of an analysis of long-wavelength observations of the Bok Globule CB244. In contrast, \citet{VPN13} conclude that any correlation is too small to evaluate, on the basis of their analysis of long-wavelength observations starless cold clumps.

Third, the isothermal simplification in Eqn. \ref{eqn:greybody} ignores that fact that dense, externally illuminated prestellar cores have negative radial temperature gradients \citep[e.g.][]{RAP13}. \citet{MJC11} have shown that temperature variations along the line of sight tend to result in the associated mass being underestimated \citep[see also][]{MWL15}. This suggests that the the probability that Oph-B11 has a mass greater than the hydrogen-burning limit may be even higher than estimated here.

\subsection{Interim summary}

We conclude that Oph-B11 probably has a mass below the hydrogen-burning limit, but probably not quite as low as estimated by \citet{AWG12}. If we accept this conclusion, it is appropriate to consider whether Oph-B11 is necessarily destined to collapse and form a brown dwarf.

\section{Could Oph-B11 be a transient condensation?}\label{SEC:ALTS}

An alternative explanation for Oph-B11 is that it is a transient condensation. In the turbulent fragmentation paradigm \citep[e.g.][]{PN02,HC08,HC09}, for every condensation with mass below the peak of the CMF ($\sim\!{\rm M}_{_\odot}\!$) that collapses to form a star there are several {\it bouncing} condensations that form and then disperse. These bouncing condensations typically spend even longer near their maximum density, with low velocity dispersion, than do the collapsing ones, so they are more likely to be seen. Moreover, as one proceeds to lower masses, bouncing condensations outnumber collapsing ones by an increasingly large factor ($\ga 30$ for condensations with $M\sim 0.030\,{\rm M}_{_\odot}$; {\AA} Nordlund, private communication).

To pass itself off as a collapsing core (that is, one with $\alpha_{_{\rm VIR}}\!=\!3R\sigma^2/GM_{_{\rm OBS}}\la 1$, where $\sigma$ is the total one-dimensional, i.e. radial, velocity dispersion), a bouncing core must probably be prolate, with an aspect ratio $q\ga \alpha^{-1}$. In addition, its long axis must be oriented close to the line of sight, $\theta\la\sin^{-1}(q^{-1})$, in order to conceal its elongation from the observer.

The first requirement is probably easily met, since turbulence generates low-dimensional structures (sheets and filaments) routinely. Bouncing condensations are in fact very unlikely to be even approximately spherical.

Assuming random orientation, the second requirement reduces to a probability of order $P\sim 1-(1-q^{-2})^{1/2}$.

\citet{AWG12} derive $\alpha\ga 0.5$, so we only need an aspect ratio $q\ga 2$, in which case the constraint on the orientation would be $P\la 0.13$.

Combining (i) the high abundance of bouncing condensations relative to collapsing ones ($\ga 30$ for a core of mass $\sim 0.030\,{\rm M}_{_\odot}$), (ii) the modest constraint on the aspect ratio ($\ga 2$) required to render Oph-B11 bouncing rather than collapsing, and (iii) the substantial likelihood ($P\la 0.13$) that the orientation of Oph-B11 renders it approximately spherical in projection, taken together, suggest that Oph-B11 could easily be a transient bouncing condensation. This does not mean that Oph-B11 is not a collapsing condensation, destined to form an isolated brown dwarf, just that searches for such objects can be expected to turn up many false positives.

There are three further considerations worth mentioning. First, at the limits of telescope sensitivity, an approximately end-on prolate condensation will be easier to detect, because it presents a greater column-density. Second, an approximately end-on bouncing prolate condensation probably presents a low line-of-sight velocity dispersion, because its longitudinal velocity dispersion is likely to be small, and the radial motions due to its contraction and expansion have small components along the line of sight. Third, the somewhat reduced density in a prolate bouncing condensation is still more than sufficient to deliver the observed $\mathrm{N}_2\mathrm{H}^+$ emission \citep[see][]{Sh15,HV16}.

\changes{The expectation of a large number of bouncing condensations is consistent with recent observations of Ophiuchus \citep{PWK15}. Here, very low mass cores ($M\lesssim0.075\,\mathrm{M_\odot}$) are common and account for at least a third of the population. However, we note that these objects are unresolved and it is possible that they are sufficiently dense to collapse.}

\section{Conclusions}\label{SEC:CONC}

This paper is concerned with the question of whether isolated brown dwarfs are likely to form by turbulent fragmentation, and in particular whether the Oph-B11 core reported by \citet{AWG12} is destined to form an isolated brown dwarf. The paper comprises three sections.

In Section \ref{SEC:SIMS}, we have attempted to constrain the parameters of turbulent flows that lead to the formation of isolated brown dwarfs, by performing well resolved Smoothed Particle Hydrodynamics simulations of core formation. These simulations use Initial Conditions in which the velocity field is characterised by a velocity magnitude, $v_{_{\rm O}}$, and a parameter $c$ which measures the extent to which the velocity field converges from all directions. We treat cores with masses of $0.020\,{\rm M}_{_\odot}$ and $0.060\,{\rm M}_{_\odot}$. A brown dwarf is assumed to have formed if a sink particle is formed. We find that brown dwarfs only form when $v_{_{\rm O}}$ and $c$ are sufficiently large. In particular the large $c$-values required suggest that brown dwarfs only form when the initial flow converges from all sides, on a scale of $\sim 0.02\,{\rm pc}$. Such flows presumably occur occasionally -- and the Oph-B11 may be a case in point -- but it seems unlikely that they are sufficiently common to make a major contribution to the observed population of isolated brown dwarfs.

In Section \ref{SEC:MASS}, we have re-analysed the observations of Oph-B11, taking uncertainties in the distance ($D$), temperature ($T$), absolute mass opacity coefficient ($\kappa_{_{230\,\mathrm{GHz}}}$) and emissivity index ($\beta$) into account. We find that the mass of Oph-B11 is likely to be below the hydrogen-burning limit, but the expectation value for the mass is probably somewhat ($\la 50\%$) higher than the range estimated by \citet{AWG12} -- {\it and} there is a significant likelihood ($14\;{\rm to}\;34\%$) that the mass of Oph-B11 is above the hydrogen burning limit. Therefore, {\it} if Oph-B11 collapses to form a star, and allowing that the process is unlikely to be $100\%$ efficient, this star will almost certainly be below the hydrogen-burning limit, i.e. a brown dwarf.

In Section \ref{SEC:ALTS}, we have considered alternative interpretations of Oph-B11 -- and other similar objects that have been, or will be, detected. We argue that a significant fraction of such objects are likely to be transient bouncing condensations, formed by turbulent flows, having prolate geometry but oriented in such a way as to conceal this, and with a low line-of-sight velocity dispersion because, with this alignment, the motions forming and/or dispersing them are largely perpendicular to the line of sight.

In summary, turbulent fragmentation may not be able to form all, or even a significant fraction of, isolated brown dwarfs, and if this is the case there is a role for an alternative formation mechanism. 

\section*{Acknowledgements}%

OL and APW gratefully acknowledge the support of a consolidated grant from the UK Science and Technology Facilities Council (ref. ST/K00926/1). DAH acknowledges the the support of the Deutsche Forschungs Gemeinschaft {\it cluster of excellence} `Origin and Structure of the Universe'. We also thank the considerate and constructive comments from the referee. The computations were performed on the Cardiff University Advanced Research Computing facility, {\sc arcca}.

\bibliographystyle{mn2e}
\bibliography{refs}

\label{lastpage}
\end{document}